\documentclass[journal]{IEEEtran}
\usepackage{cite}
\usepackage{amsmath,amssymb,amsfonts}
\usepackage{float}
\usepackage{graphicx}
\usepackage{textcomp}
\usepackage{xcolor}
\usepackage{algorithm}
\usepackage{algpseudocode}
\usepackage{booktabs}
\usepackage{multirow}
\usepackage{bm}
\usepackage{url}
\usepackage{braket}
\usepackage{pdfpages}
\usepackage{amsthm}

\newcommand{\tablefontsize}{\fontsize{8}{9.2}\selectfont}

\newcommand{\rev}[1]{#1}
\newcommand{\revold}[1]{#1} 

\begin{document}

\title{Optimizing Encoder Circuits of Entanglement-Assisted Quantum LDPC Codes via Beam Search}

\author{Aditya~Sodhani,~\IEEEmembership{Member,~IEEE,}
        Pavan~Kumar,
        Shayan~Srinivasa~Garani,~\IEEEmembership{Senior~Member,~IEEE,}
        and~Keshab~K.~Parhi,~\IEEEmembership{Life~Fellow,~IEEE}%
\thanks{A. Sodhani and K. K. Parhi are with the Department of Electrical
and Computer Engineering, University of Minnesota, Minneapolis, MN 55455 USA
(e-mail: sodha005@umn.edu; parhi@umn.edu).}%
\thanks{P. Kumar and S. S. Garani are with the Department of Electronic
Systems Engineering, Indian Institute of Science, Bengaluru 560012, India
(e-mail: pavankumar1@iisc.ac.in; shayangs@iisc.ac.in).}%
\vspace{-1.2em}%
}

\maketitle
\begin{abstract}
In encoder circuits built on the stabilizer formalism, the dominant contribution to circuit complexity comes from the use of controlled (CNOT) gates, making CNOT-count reduction a central circuit-design objective. Entanglement-assisted (EA) quantum QC-LDPC codes offer strong error-correction capabilities with structured parity-check matrices, but their practical use depends on efficient encoder circuits and the availability of pre-shared Bell pairs (ebits). In this paper, we adopt a prior entanglement-assisted QC-LDPC (EAQC) encoder construction. We formulate the encoder optimization as a search over $\mathrm{GF}(2)$ row operations that decompose the binary matrix derived from its CNOT sub-sequence. We solve this problem using a beam search algorithm guided by a Hamming-distance heuristic. For the tested EA quantum QC-LDPC code families, the proposed method achieves CNOT-count reductions of 7.3--34.0\% relative to the baseline EAQC encoder. \revold{The optimized circuits also outperform the Patel--Markov--Hayes and greedy cost-minimization baselines, and are verified by stabilizer-tableau simulation.} These results show that substantial encoder simplification is possible for structured EA QC-LDPC codes.
\end{abstract}

\begin{IEEEkeywords}
Circuit synthesis, CNOT optimization, beam search, quantum error correction, entanglement-assisted codes, QC-LDPC codes.
\end{IEEEkeywords}


\section{Introduction}
\label{sec:intro}

\begingroup
\setlength{\parskip}{0pt}

Quantum computers exploit superposition and entanglement for information processing, with potential applications in combinatorial optimization, machine learning, and other computationally hard problems. However, quantum states are highly sensitive to decoherence, imperfect control, and measurement noise, which makes reliable large-scale quantum computation difficult~\cite{bravyi2024high}. Quantum error-correcting codes (QECCs) based on stabilizers protect quantum information against such errors~\cite{shor1995scheme,gottesman1997stabilizer} and are realized by encoder circuits composed primarily of Hadamard and CNOT gates. Since the CNOT gate dominates the implementation cost in current quantum hardware~\cite{maslov2007linear}, reducing the CNOT count is important for encoder efficiency and circuit fidelity.

The Calderbank--Shor--Steane (CSS) construction~\cite{calderbank1996good,steane1996multiple} derives quantum codes from pairs of classical codes satisfying a symplectic orthogonality condition. Entanglement-assisted (EA) quantum codes broaden this framework by using pre-shared Bell pairs (ebits), allowing constructions from classical codes that are not dual-containing~\cite{brun2006correcting}. Quasi-cyclic low-density parity-check (QC-LDPC) codes are attractive candidates for EA construction because their parity-check matrices have regular circulant structure and favorable error-correction properties~\cite{gallager1962low,fossorier2004quasicyclic}. Recent work~\rev{\cite{kumar2026entanglement}} introduced several families of EA quantum QC-LDPC codes with guaranteed girth and minimum-distance properties. \revold{Related EA constructions based on quasi-cyclic codes have also been reported~\cite{matsui2025entanglement}.}

While code construction and decoding have been widely studied, the complexity of the corresponding encoder circuits has received less attention. Prior work has addressed both the derivation of valid encoders and the simplification of the resulting circuits. For stabilizer block codes, algebraic manipulation of stabilizer generators yields efficient encoding circuits~\cite{cleve1997efficient}. Encoder constructions for EA codes extend the stabilizer framework with shared ebits. Circuit-level simplification and gate-count optimization of stabilizer encoders have also been studied~\cite{mondal2024optimization,mondal2024quantum}.

\revold{With the non-CNOT Clifford layers held fixed, the remaining CNOT sub-sequence of the encoder can be represented by an invertible binary matrix over $\mathrm{GF}(2)$. The resulting optimization problem is a linear reversible circuit synthesis problem. Existing methods use the same problem formulation but differ in how they explore the possible decompositions. Standard Gaussian elimination follows a fixed pivot order, while PMH combines identical sub-row patterns within fixed sections before completing the elimination~\cite{markov2008optimal}. Greedy methods such as AECM and MCG score candidate row or column operations with matrix-cost functions and keep only one path at each step~\cite{schaeffer2014cost,de2021gaussian}. A recent work~\cite{webster2025heuristic} uses exact synthesis for small matrices, A* search (a heuristic best-first search) for intermediate sizes, and greedy synthesis for larger matrices. Related works have also considered canonical-form Clifford synthesis, as well as non-binary encoder optimization~\cite{aaronson2004improved,sodhani2026encoder}. Reinforcement-learning approaches to CNOT synthesis have also been proposed~\cite{romanello2025cnot,cossio2026alphacnot}.}

\revold{The proposed method differs in its search policy, retaining up to $w$ competing decompositions of the whole encoder matrix ($M$). It is not tied to a fixed pivot order as in Gaussian elimination, to the fixed matrix sections used by PMH~\cite{markov2008optimal}, or to the original ordering of the entanglement-assisted QC-LDPC (EAQC) encoder layers~\cite{sharma2025encoding}. It can therefore retain a locally weaker operation when that operation enables a larger reduction at a later step. In this sense, it lies between single-path greedy synthesis and exact search, exploring more alternatives than a greedy method while restricting the search far more aggressively than exact or A* methods.}

\revold{The contribution of this work is an approach based on beam search~\cite{ow1988filtered} that keeps a limited number of candidate decompositions and optimizes the EAQC-derived encoder matrix as a whole. Table~\ref{tab:notation} summarizes the notation used throughout. We combine the complete CNOT sub-sequence of $\mathcal{E}=\mathcal{D}^{\dagger}$ into a single matrix $M$, while keeping the non-CNOT Clifford layers and ebit structure unchanged. Candidate states are ranked using $h(M')=\mathrm{wt}(M'\oplus I)$. This heuristic favors row operations that remove several mismatches with the identity at once. When $w=1$, the method reduces to a single-path greedy search under this heuristic. When $w>1$, it keeps several partial decompositions and can explore different operation orders. This additional search flexibility comes with increased runtime and memory use. Moreover, pruning to $w$ states removes any guarantee of finding a minimum-CNOT circuit, and the result depends on the beam width, the heuristic, and the row structure of $M$. The method is therefore most useful when the encoder matrix contains strong row correlations. We evaluate it on the \rev{Theorem-5} EA quantum QC-LDPC code family \rev{and the Theorem-6 and Theorem-7 families} in~\rev{\cite{kumar2026entanglement}}, together with two additional small EA codes, and compare it with the EAQC, PMH, and greedy baselines.}

This paper is organized as follows. Section~\ref{sec:prelim} presents the encoder circuit model. Section~\ref{sec:beam} presents the proposed method. Section~\ref{sec:results} presents the results. Section~\ref{sec:conclusion} concludes the paper.

\endgroup

\begin{figure}[t]
\centering
\includegraphics[width=\columnwidth]{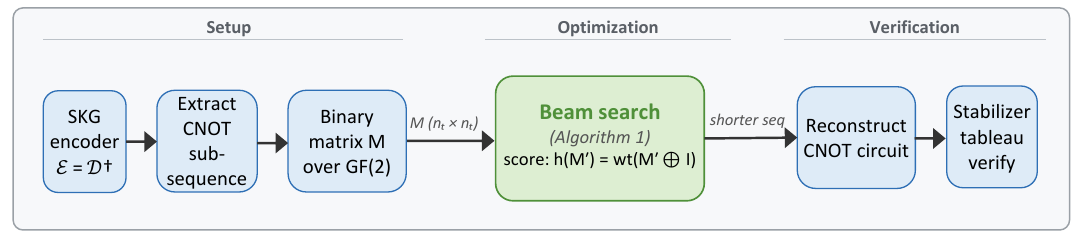}
\caption{Architecture of the proposed CNOT-count optimization flow.}
\label{fig:method_flow}
\end{figure}

\section{Encoder Circuit Model}
\label{sec:prelim}

An $[[n,k,d;c]]$ EA quantum code encodes $k$ logical qubits into $n$ physical qubits using $c$ pre-shared Bell pairs (ebits), so the encoder acts on $n_t = n + c$ qubits. \revold{The parity-check matrices $H_x$ ($X$-type) and $H_z$ ($Z$-type) are augmented with ebit columns to form the extended matrices $H_{ex}$ and $H_{ez}$, which satisfy $H_{ex}H_{ez}^{T}\equiv 0\pmod{2}$.}

Following the EAQC encoder construction~\cite{sharma2025encoding}, we build a Clifford decoder $\mathcal{D} = \mathcal{D}_5\mathcal{D}_4\mathcal{D}_3\mathcal{D}_2\mathcal{D}_1\mathcal{D}_0$, where $\mathcal{D}_5$ is a Hadamard layer and $\mathcal{D}_0$--$\mathcal{D}_4$ are CNOT layers; the encoder $\mathcal{E} = \mathcal{D}^{\dagger}$ runs the six layers in reverse. The Supplementary Material (Sections~S2 and~S3) gives the per-layer description of $\mathcal{D}_0$--$\mathcal{D}_5$, the block-role glossary, and the QC-LDPC parity-check construction (\rev{Theorem~5 of~\cite{kumar2026entanglement}}) that defines the code family we optimize.

\begin{table}[t]
\centering
\caption{Notation summary.}
\label{tab:notation}
\tablefontsize
\setlength{\tabcolsep}{4.0pt}
\renewcommand{\arraystretch}{1.08}
\begin{tabular}{|l|l|}
\hline
\textbf{Symbol} & \textbf{Meaning} \\
\hline
$[[n,k,d;c]]$ & length, logical qubits, distance, ebits \\
\hline
$n_t = n + c$ & total qubits in the encoder \\
\hline
$H_x$, $H_z$ & X-/Z-type parity-check matrices \\
\hline
$D_x$, $D_z$ & ebit columns, $D_x D_z^T = H_x H_z^T$ \\
\hline
$H_{ex}$, $H_{ez}$ & extended matrices $[H_x \mid D_x]$, $[H_z \mid D_z]$ \\
\hline
$p$, $\ell$ & circulant size; block rows per matrix \\
\hline
$r_B$ & $\mathrm{rank}(H_x H_z^T)$ \\
\hline
$\mathcal{D}$, $\mathcal{E}$ & EAQC Clifford decoder; encoder $\mathcal{E}=\mathcal{D}^{\dagger}$ \\
\hline
$M$ & encoder matrix; $\mathrm{GF}(2)$ matrix of CNOT sub-sequence \\
\hline
$L_{\mathrm{EAQC}}$ & CNOT count of the baseline EAQC encoder \\
\hline
$h(\cdot)$ & Hamming heuristic $\mathrm{wt}(\cdot \oplus I)$ \\
\hline
$w$, $d_{\max}$ & beam width; maximum search depth \\
\hline
\end{tabular}
\end{table}

\subsection{Encoder Matrix Representation}
\label{sec:encoder}

The CNOT sub-sequence of the encoder $\mathcal{E}$ acts on $n_t$ qubits. Starting from the identity, applying CNOTs in order produces an invertible binary matrix $M\in\mathrm{GL}(n_t,\mathrm{GF}(2))$ on the qubit-value register, which we call the \emph{encoder matrix}: each CNOT$(j\to i)$ realizes the row update $M[i]\leftarrow M[i]\oplus M[j]$. The Hadamard layer ($\mathcal{D}_5$) acts only on the X-ancilla register and is invariant under our optimization. Let $L_{\text{EAQC}}$ denote the number of CNOTs derived by the EAQC encoder construction; this sequence is a valid decomposition of $M$ into elementary row-XOR operations (transvections). Since transvections are self-inverse, our objective is equivalent to finding a shorter sequence of row-XOR operations that reduces $M$ to the identity.

\section{Proposed Encoder Optimization}
\label{sec:beam}

\subsection{Problem Formulation}

Given the encoder matrix $M \in \mathrm{GF}(2)^{n_t \times n_t}$, we seek a minimum-length sequence of row operations $(i_1, j_1), \ldots, (i_L, j_L)$, each applying $R_{i_t} \leftarrow R_{i_t} \oplus R_{j_t}$, that transforms $M$ into the identity $I$. Since CNOT gates are self-inverse, each operation corresponds to one CNOT with control $j_t$ and target $i_t$, so $L$ is the optimized CNOT count and the problem is a minimum-factor decomposition of $M$ into elementary row-addition matrices. The search space has $n_t(n_t - 1)$ possible operations per step.

\textit{Circuit equivalence.} Different CNOT sequences can implement the same encoder matrix $M$; the shortest is what our optimization targets. \revold{A step-by-step four-qubit example, in which a 5-CNOT sequence and a 3-CNOT sequence realize the same $M$, is given in the Supplementary Material (Section~S5).} The Algorithm~\ref{alg:beam_search}, based on beam search, performs the analogous comparison over a much larger search space: on $[[9,4;1]]$ it discovers a 13-CNOT sequence whose cumulative matrix matches that of the 17-CNOT EAQC sequence.

\subsection{Beam Search Algorithm}

Fig.~\ref{fig:method_flow} summarizes the end-to-end optimization flow, from extracting the EAQC CNOT sub-sequence and forming the binary matrix $M$, to beam-search optimization, CNOT reconstruction, and stabilizer-tableau verification. We define the Hamming distance heuristic
\begin{equation}
h(M') = \text{wt}(M' \oplus I),
\label{eq:heuristic}
\end{equation}
which counts the number of positions where the current matrix $M'$ differs from the identity. This measures how far the current state is from the goal and serves as the evaluation function in our beam search. \revold{The heuristic satisfies $h(M')=0$ if and only if $M'=I$. Since each CNOT changes one row, the change in $h(M')$ depends on the overlap between the source row and the entries in the target row that differ from the identity. Ranking candidates by lower $h(M')$, therefore, favors operations that remove several mismatches at once. This effect is quantified by the clearing-rate analysis in the Supplementary Material (Section~S7).}

The beam search proceeds iteratively over depth levels $d = 1, 2, \ldots, d_{\max}$, where $d_{\max} = L_{\text{EAQC}}$. This bound is valid because the reversed EAQC sequence already achieves $L_{\text{EAQC}}$ row operations, so the optimum cannot exceed it. At each depth, all $n_t(n_t - 1)$ possible row operations are applied to every state in the current beam $\mathcal{B}$, producing a candidate set $\mathcal{C}$. Each candidate is scored by $h(\cdot)$, and only the top-$w$ states with lowest Hamming distance are retained for the next depth level. If any candidate equals the identity matrix $I$, the search terminates and returns the corresponding operation sequence. The beam is initialized with $\mathcal{B} = \{M\}$.

\begin{algorithm}[t]
\caption{Beam Search for Encoder CNOT Reduction}
\label{alg:beam_search}
\begin{algorithmic}[1]
\Require Encoder matrix $M$, beam width $w$, maximum depth $d_{\max} = L_{\text{EAQC}}$
\Ensure Operation sequence $\mathcal{S}$ such that $M \rightarrow I$, if found
\State $\mathcal{B} \gets \{(M, \emptyset)\}$
\For{$d = 1$ to $d_{\max}$}
    \State $\mathcal{C} \gets \emptyset$
    \ForAll{$(M', \mathcal{S}') \in \mathcal{B}$}
        \ForAll{ordered pairs $(i,j)$ with $i \neq j$}
            \State $\widetilde{M} \gets M'$ with row update $R_i \leftarrow R_i \oplus R_j$
            \State $\widetilde{\mathcal{S}} \gets \mathcal{S}'$ appended with $(i,j)$
            \If{$\widetilde{M} = I$}
                \State \Return $\widetilde{\mathcal{S}}$
            \EndIf
            \State score $\gets h(\widetilde{M})$
            \State add $(\widetilde{M}, \widetilde{\mathcal{S}}, \text{score})$ to $\mathcal{C}$
        \EndFor
    \EndFor
    \State retain the top-$w$ candidates from $\mathcal{C}$ with the lowest scores as the new beam $\mathcal{B}$
\EndFor
\State \Return failure
\end{algorithmic}
\end{algorithm}

Algorithm~\ref{alg:beam_search} summarizes the proposed search procedure. \rev{Ties among equal-$h(M')$ candidates and duplicate states are resolved by fixed deterministic rules, stated with a worked example in the Supplementary Material (Section~S6), so repeated runs produce identical results.} \revold{Beam search retains at most $w$ states at each depth, so the partial sequences that would reach an optimal decomposition may be discarded along the way. The method is therefore heuristic and does not guarantee a minimum-CNOT decomposition. This limitation is not specific to beam search: the greedy cost-minimization and reinforcement-learning methods~\cite{schaeffer2014cost,de2021gaussian,romanello2025cnot,cossio2026alphacnot} are likewise heuristic. Only exact synthesis~\cite{webster2025heuristic} can guarantee a minimum-CNOT circuit. Finding this minimum among all circuit-equivalent decompositions is a combinatorial optimization problem that does not scale to our largest instances, which reach $n_t=122$. The proposed beam search is therefore a pragmatic tradeoff between the algebraic EAQC encoder construction and this unscalable optimum.}

\subsection{Baselines}

We compare the proposed beam-search optimizer with \revold{three} baselines. The first is the EAQC encoder, whose CNOT count is $L_{\mathrm{EAQC}}$. This encoder is a valid reference for the EA quantum codes considered here because it preserves the ebit-pair stabilizer structure and the fixed Hadamard layer on the $X$-stabilizer ancillas. Applying Gaussian elimination directly to the classical parity-check matrices does not, in general, preserve either property. The second is the Patel--Markov--Hayes (PMH) algorithm~\cite{markov2008optimal}, applied to the binary matrix $M$ induced by the EAQC CNOT sub-sequence. PMH performs section-wise row reduction with sub-row pattern matching and achieves asymptotic CNOT count $O(n_t^2/\log n_t)$. \revold{The third is a greedy cost-minimization baseline that follows the AECM and MCG methods~\cite{schaeffer2014cost,de2021gaussian}. At each step it applies the row operation that gives the largest reduction in $h(M')$, using a two-step lookahead to escape flat regions, and commits to that single path.}
\vspace{-0.8em}
\subsection{\revold{Runtime and Memory}}

\revold{At each search depth, Algorithm~\ref{alg:beam_search} evaluates at most $w n_t(n_t-1) = O(w n_t^2)$ candidates. Over at most $d_{\max}$ depths, this gives $O(w n_t^2 d_{\max})$ candidate evaluations. Since applying a row XOR and computing the heuristic require $O(n_t)$ word operations per candidate, the total computational cost is $O(w n_t^3 d_{\max})$ word operations. In practice, the search converges well before $d_{\max}$ for codes with exploitable structure.}

\revold{Table~\ref{tab:width_runtime} reports the runtime for each tested code and beam width. The experiments were run using a single-threaded Python/NumPy program on a computer with an Intel Core Ultra~7 155H processor \rev{(up to 4.8\,GHz)} and 32~GB of RAM. Memory use remained below a few hundred megabytes for all tested cases.}

\section{Results and Discussion}
\label{sec:results}

We evaluate the proposed beam-search optimizer on \rev{ten} EA QC-LDPC codes from the \rev{Theorem-5, Theorem-6, and Theorem-7 constructions of~\cite{kumar2026entanglement}}, restated in the Supplementary Material (Section~S3). Here $p$ denotes the circulant size and $\ell$ the per-matrix block row count, with larger $\ell$ producing a denser encoder matrix. To include shorter block-length examples, we also evaluate two additional small EA benchmark codes, $[[6,2;2]]$ and $[[8,2;2]]$. \rev{Their small size keeps the full search trajectory traceable, $[[6,2;2]]$ serves as the step-by-step worked example in the Supplementary Material (Section~S6).} Beam widths from $w=1$ to $w=500$ are tested depending on the code size, and the smallest verified CNOT count is reported for each code. \revold{In addition to the EAQC and PMH baselines, the evaluation includes the greedy cost-minimization baseline of Section~III-C.} Each optimized circuit passes three checks: the row operations reduce the extracted matrix to $I$; the forward CNOT sequence reconstructs $M$ from the identity; and the full encoder (fixed Hadamard layer plus optimized CNOTs) passes the stabilizer-tableau test~\cite{aaronson2004improved}. This test verifies that the output generators span the target EA-CSS stabilizer group. All \rev{twelve} circuits pass.
\vspace{-0.6em}

\subsection{Gate-Count Comparison}

Table~\ref{tab:gate_results} presents the CNOT gate counts and gate-reduction percentages across all methods: the ``EAQC'' column is the construction count $L_{\text{EAQC}}$, ``PMH'' is the Patel--Markov--Hayes~\cite{markov2008optimal} result, \revold{``Greedy'' is the cost-minimization baseline described in Section~III-C,} ``Beam'' is the best beam-search result over the tested widths, and the reduction is measured relative to the baseline EAQC encoder.

\begin{table}[t]
\centering
\caption{Gate-count comparison for EA QC-LDPC encoder circuits. \revold{$\Delta = \mathrm{EAQC} - \mathrm{Beam}$.} \rev{``n.c.'': no complete decomposition (Section~IV-B). For Theorem-6/7 rows, $\ell$ is the total block-row count.}}
\label{tab:gate_results}
\tablefontsize
\setlength{\tabcolsep}{2.6pt}
\renewcommand{\arraystretch}{1.08}
\begin{tabular}{|l|c|c|c|c|c|c|c|c|c|}
\hline
\textbf{Code} & $p$ & $\ell$ & $n_t$ & \textbf{EAQC} & \textbf{PMH} & \revold{\textbf{Greedy}} & \textbf{Beam} & \revold{$\bm{\Delta}$} & \textbf{Red.} \\
\hline
$[[9,4;1]]$        &  3 & 1 &  10 &  17 &  16 & \revold{14} & \textbf{13}  & \revold{4}   & 23.5\% \\
\hline
$[[25,16;1]]$      &  5 & 1 &  26 &  38 &  36 & \revold{34} & \textbf{33}  & \revold{5}   & 13.2\% \\
\hline
$[[25,8;1]]$       &  5 & 2 &  26 &  89 &  89 & \revold{66} & \rev{\textbf{61}}  & \rev{28}  & \rev{31.5\%} \\
\hline
$[[49,36;1]]$      &  7 & 1 &  50 &  82 &  78 & \revold{74} & \textbf{74}  & \revold{8}   &  9.8\% \\
\hline
$[[49,12;1]]$      &  7 & 3 &  50 & 338 & 340 & \revold{233} & \textbf{223} & \revold{115} & 34.0\% \\
\hline
$[[121,100;1]]$    & 11 & 1 & 122 & 218 & 210 & \revold{202} & \textbf{202} & \revold{16}  &  7.3\% \\
\hline
\rev{$[[20,10;4]]$} & \rev{5} & \rev{3} & \rev{24} & \rev{47} & \rev{51} & \rev{n.c.} & \rev{\textbf{37}} & \rev{10} & \rev{21.3\%} \\
\hline
\rev{$[[25,16;9]]$} & \rev{5} & \rev{2} & \rev{34} & \rev{179} & \rev{286} & \rev{n.c.} & \rev{\textbf{156}} & \rev{23} & \rev{12.8\%} \\
\hline
\rev{$[[42,10;6]]$} & \rev{7} & \rev{6} & \rev{48} & \rev{281} & \rev{283} & \rev{n.c.} & \rev{\textbf{192}} & \rev{89} & \rev{31.7\%} \\
\hline
\rev{$[[42,22;6]]$} & \rev{7} & \rev{4} & \rev{48} & \rev{367} & \rev{375} & \rev{n.c.} & \rev{\textbf{246}} & \rev{121} & \rev{33.0\%} \\
\hline
$[[6,2;2]]$      & -- & -- &   8 &  11 &  18 & \revold{10} & \textbf{9}   & \revold{2}   & 18.2\% \\
\hline
$[[8,2;2]]$        & -- & -- &  10 &  13 &  13 & \revold{10} & \textbf{10}  & \revold{3}   & 23.1\% \\
\hline
\end{tabular}
\end{table}

Several observations emerge from Table~\ref{tab:gate_results}. First, PMH~\cite{markov2008optimal} saves little over EAQC (\rev{at most 8 CNOTs on any code}, and is worse \rev{than the baseline} on $[[49,12;1]]$ and $[[6,2;2]]$ from the 3-CNOT row-swap overhead \rev{ and on all four Theorem-6/7 codes}): the baseline cost is governed by the row structure of $M$, not by the elimination ordering. Beam search improves on it by exploiting overlaps among the rows of $M$. Second, beam search reduces the count on every code, from 7.3\% on $[[121,100;1]]$ to 34.0\% on $[[49,12;1]]$\rev{, while the largest absolute saving of 121 CNOTs occurs on the Theorem-6 code $[[42,22;6]]$}. Third, \rev{among the tested codes,} denser codes (larger $\ell$) admit larger relative reductions, since the denser block structure creates correlated, overlapping rows that beam search exploits via row sharing. \revold{Fourth, the greedy baseline is fast \rev{(within 5\,s on the machine of Section~III-D) on each of the eight codes it completes}. \rev{However, compared with the proposed beam-search approach, its performance is the same on} the two sparsest $\ell=1$ instances and on $[[8,2;2]]$, but is strictly worse on the other five\rev{, and it fails to complete on the four Theorem-6/7 codes}. In absolute terms, the beam search optimization saves \rev{424} CNOTs across the \rev{twelve} codes (column $\Delta$ of Table~\ref{tab:gate_results}).} \rev{For the smallest instance $[[6,2;2]]$, an exhaustive search confirms that no circuit with eight or fewer CNOTs implements the encoder, so the 9-CNOT beam-search circuit is optimal. The same check is already infeasible at $n_t=10$, so optimality can be proven only for the smallest instance.}

\begin{figure}[t]
\centering
\includegraphics[width=0.88\columnwidth]{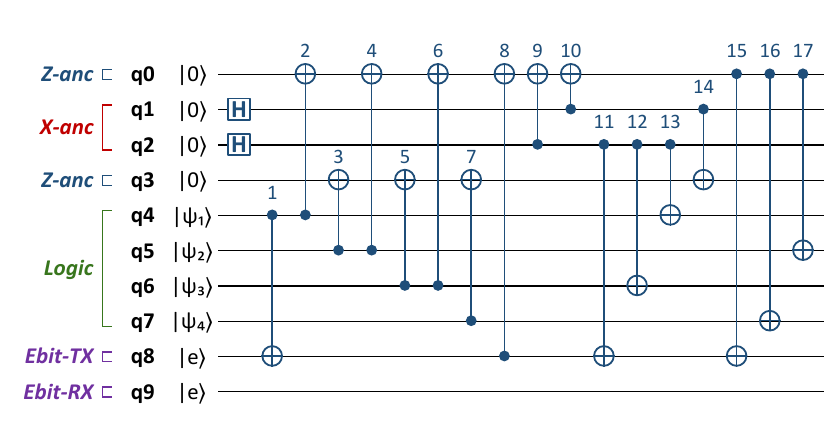}\\[2pt]
\includegraphics[width=0.88\columnwidth]{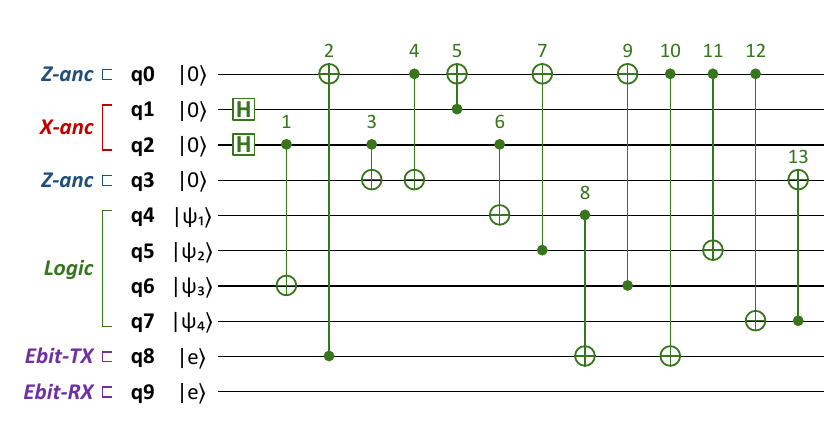}
\vspace{-0.6em}
\caption{\rev{Encoder circuits for $[[9,4;1]]$. Top: baseline EAQC encoder, 2 H and 17 CNOTs. Bottom: beam-search-optimized encoder, 2 H and 13 CNOTs.}}
\label{fig:circuits_941}
\end{figure}

To illustrate the optimization concretely, \rev{Fig.~\ref{fig:circuits_941} shows} the complete CNOT encoder circuits for the $[[9,4;1]]$ code ($p = 3$, $\ell = 1$, $n_t = 10$) under the EAQC encoder construction and beam search, respectively. The 10~qubits comprise two X-stabilizer ancillas ($q_1$, $q_2$), two Z-stabilizer ancillas ($q_0$, $q_3$), four logical qubits ($q_4$--$q_7$), and one pre-shared ebit pair ($q_8$ at the sender, $q_9$ at the receiver). The EAQC encoder applies 17~CNOTs from $\mathcal{D}_1$--$\mathcal{D}_4$ in reverse order, while the beam search discovers a 13-operation sequence: operations that reduce $h(M')$ by more than one are preferred by the beam selection, leading to shorter overall sequences.

\vspace{-0.9em}
\subsection{Beam Width Sensitivity}

\begin{table}[t]
\centering
\caption{CNOT count and runtime versus beam width $w$; each entry is the count with runtime \rev{($\mathrm{s}$)} in parentheses. Per-run time caps range from \rev{120\,$\mathrm{s}$} ($n_t \le 10$) to \rev{1200\,$\mathrm{s}$} ($n_t = 122$). ``--'': not run; ``n.c.'': no complete decomposition \rev{(time cap reached or depth limit exhausted)}. \rev{The sweep is restricted to the listed widths and time caps.}}
\label{tab:width_runtime}
\fontsize{7}{8.4}\selectfont
\setlength{\tabcolsep}{2.2pt}
\renewcommand{\arraystretch}{1.08}
\begin{tabular}{|l|c|c|c|c|c|c|}
\hline
\textbf{Code} & $n_t$ & $w=1$ & $w=10$ & $w=50$ & $w=100$ & $w=500$ \\
\hline
$[[6,2;2]]$      &   8 & 10 ($<$0.1)  & 9 ($<$0.1)    & 9 (0.5)    & 9 (0.4)    & 9 (2.2) \\
\hline
$[[8,2;2]]$      &  10 & 10 ($<$0.1)  & 10 ($<$0.1)   & 10 (0.4)   & 10 (0.6)   & 10 (9.5) \\
\hline
$[[9,4;1]]$      &  10 & 14 ($<$0.1)  & 14 ($<$0.1)   & 14 (0.4)   & 14 (0.8)   & 13 (18.5) \\
\hline
$[[25,16;1]]$    &  26 & 34 (0.2)  & 34 (11.4)  & 33 (72.2)  & 33 (153.4) & -- \\
\hline
$[[25,8;1]]$     &  26 & 66 (0.4)  & 63 (26.6)  & 62 (152.5) & 61 (319.7) & -- \\
\hline
$[[49,36;1]]$    &  50 & 74 (19.1) & 74 (200.5) & n.c. (901.9) & -- & -- \\
\hline
$[[49,12;1]]$    &  50 & 233 (78.3) & 223 (619.9) & n.c. (902.9) & -- & -- \\
\hline
$[[121,100;1]]$  & 122 & 202 (601.4) & n.c. (1224.1) & -- & -- & -- \\
\hline
\rev{$[[20,10;4]]$} & \rev{24} & \rev{45 (0.1)} & \rev{44 (1.6)} & \rev{37 (9.0)} & \rev{37 (20.3)} & \rev{--} \\
\hline
\rev{$[[25,16;9]]$} & \rev{34} & \rev{n.c. (3.6)} & \rev{n.c. (47.4)} & \rev{156 (81.6)} & \rev{n.c. (530.5)} & \rev{--} \\
\hline
\rev{$[[42,10;6]]$} & \rev{48} & \rev{n.c. (15.8)} & \rev{n.c. (165.9)} & \rev{192 (440.5)} & \rev{--} & \rev{--} \\
\hline
\rev{$[[42,22;6]]$} & \rev{48} & \rev{n.c. (28.7)} & \rev{n.c. (289.0)} & \rev{246 (551.0)} & \rev{--} & \rev{--} \\
\hline
\end{tabular}
\end{table}
\vspace{-0.2em}

\revold{Beam width $w$ controls the trade-off between CNOT count and runtime. Table~\ref{tab:width_runtime} shows that larger widths are more useful for the denser codes. For $[[25,8;1]]$, the CNOT count decreases from 66 at $w=1$ to 61 at $w=100$, while for $[[49,12;1]]$, it decreases from 233 at $w=1$ to 223 at $w=10$. In contrast, the sparse $\ell=1$ codes $[[49,36;1]]$ and $[[121,100;1]]$ already reach their reported counts at $w=1$. Runtime increases quickly with $w$: the $n_t=50$ cases do not complete at $w=50$, and the $n_t=122$ case does not complete at $w=10$, within the stated time limits. These results support the use of small or moderate beam widths for larger matrices.} \rev{For larger encoder matrices, the method therefore operates at small beam widths in practice, and the gain over $w=1$ is governed by the row structure. The three largest Theorem-6/7 codes complete only at $w=50$ (Table~\ref{tab:width_runtime}).}
\vspace{-0.5em}
\subsection{Circuit Depth}

Although the proposed objective is CNOT-count reduction, we also estimate the two-qubit depth of the optimized CNOT sub-sequences using as-soon-as-possible (ASAP) scheduling. The scheduler (Supplementary Material, Section~S4, Algorithm~1) walks the forward CNOT list in its given time order and assigns each gate to the earliest time step in which both of its qubits are free, so two CNOTs are placed in the same time step if and only if their qubit sets are disjoint.

\begin{figure}[t]
\centering
\includegraphics[width=0.68\columnwidth]{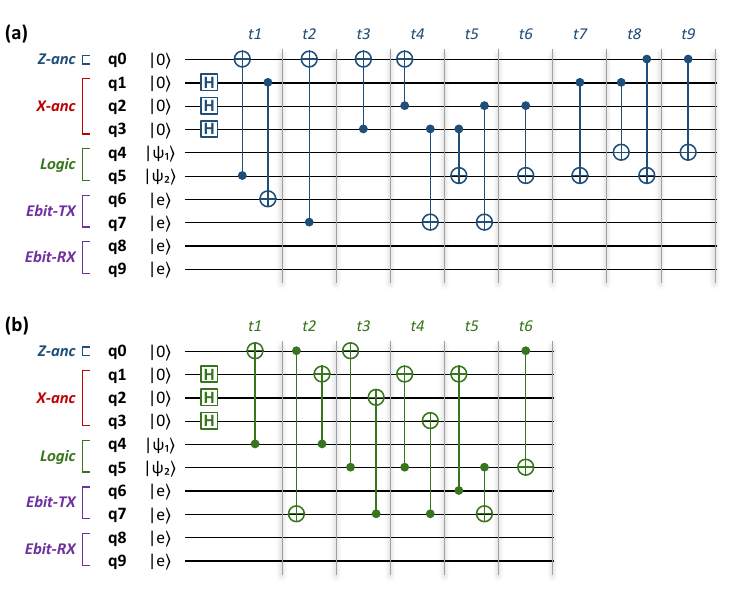}
\caption{ASAP-scheduled encoder circuits for $[[8,2;2]]$: (a) baseline EAQC, 13 CNOTs in 9 time steps; (b) beam-optimized, 10 CNOTs in
6 time steps. CNOTs acting on disjoint qubits share a time step
(dashed columns).}
\label{fig:depth_822}
\end{figure}

\revold{The optimized circuits show mixed count--depth behavior. For the $[[8,2;2]]$ code, the ASAP two-qubit depth decreases from 9 for the baseline EAQC circuit to 6 for the beam-search circuit (Fig.~\ref{fig:depth_822}). For the $[[25,8;1]]$ code, it decreases from 26 to \rev{22}; the corresponding circuits are provided in the Supplementary Material (Section~S8). For the remaining codes, the depth is preserved or slightly increased, since the objective minimizes CNOT count alone and does not account for two-qubit depth. Our recent work~\cite{sodhani2026noise} studies alternative heuristics and the joint optimization of CNOT count, two-qubit depth, and noise within the same synthesis framework.}
\vspace{-0.8em}
\section{Conclusion}
\label{sec:conclusion}

We presented a beam search method for reducing CNOT counts in EA QC-LDPC encoder circuits\rev{, achieving} $7.3$--$34.0\%$ CNOT reduction across all \rev{twelve} tested codes, each verified by stabilizer-tableau simulation to preserve the encoded code-space. The reduction is controlled primarily by the block-row structure of the encoder matrix \rev{for the tested code family}, with denser ($\ell \geq 2$) codes admitting larger gains. A natural next direction is depth-aware optimization that jointly minimizes CNOT count and two-qubit depth, and ultimately incorporates hardware connectivity~\cite{mondal2024optimized}. \revold{Error propagation, hook errors, and circuit-level logical failure are addressed in our recent work~\cite{sodhani2026noise}, while compatibility with fault-tolerant syndrome extraction is left for future work. Beyond the depth- and connectivity-aware objectives named above, natural extensions include noise-aware scoring functions and evaluation on further EA code constructions.}

\newpage


\section*{Supplementary material}

# Supplementary Material: Optimizing Encoder Circuits of Entanglement-Assisted Quantum LDPC Codes via Beam Search

Aditya Sodhani, *Member, IEEE*; Pavan Kumar; Shayan Srinivasa Garani, *Senior Member, IEEE*; and Keshab K. Parhi, *Life Fellow, IEEE*

## S1 Introduction

This supplement contains the baseline-construction material referenced in the main paper: the full per-layer description of the entanglement-assisted QC-LDPC (EAQC) decoder, the QC-LDPC parity-check construction and the code-family theorems, a four-qubit circuit-equivalence example, a step-by-step worked example of the beam search on the $[[6, 2; 2]]$ code, the clearing-rate analysis of the block-row density effect, and the baseline EAQC and beam-search-optimized CNOT encoder circuits for $[[25, 8; 1]]$, whose optimized two-qubit depth 22 is strictly smaller than that of the baseline EAQC encoder 26. The circuits below are the sequences whose as soon as possible (ASAP) scheduling yields those depths; both have been verified by stabilizer-tableau simulation.

The figures use the same drawing conventions as Fig. 2 of the main paper: horizontal qubit wires labeled $q_i$ with the initial states immediately to the right, color-coded brackets identifying the post-decoder roles (Z-stabilizer ancilla, X-stabilizer ancilla, logical qubit, sender ebit half, receiver ebit half), single-qubit Hadamards drawn as boxed 'H', and CNOTs (a filled control dot, a vertical link, and an oplus target circle).

## S2 EAQC Decoder Construction in Detail

Section II of the main paper introduces the EAQC decoder $\mathcal{D} = \mathcal{D}_5\mathcal{D}_4\mathcal{D}_3\mathcal{D}_2\mathcal{D}_1\mathcal{D}_0$ and the encoder $\mathcal{E} = \mathcal{D}^\dagger$ at a high level [1]. In overview, $\mathcal{D}_0$ is a column-permutation/row-reduction step on $H_{ex}$; $\mathcal{D}_1$–$\mathcal{D}_4$ are CNOT layers that bring $H_{ex}$ and $H_{ez}$ to a standardized pivot form; $\mathcal{D}_5$ is a Hadamard layer. Here we give the full per-layer description used in the implementation. We use the notation of §II-A of the main paper: $H_{ex} = [\, H_x \mid D_x \,]$ and $H_{ez} = [\, H_z \mid D_z \,]$ are the extended X- and Z-parity-check matrices, with the ebit columns $D_x, D_z \in \mathrm{GF}(2)^{\rho_i \times c}$ chosen so that $D_x D_z^\top = H_x H_z^\top$; $r_B$ is the rank of the Z-stabilizer support on the X-anc block (the Z-pivot block $B_1$ of $\mathcal{D}_2$ below), with $r_B \le c$.

*Block layout.* We first reorder the $n_t = n + c$ qubits into four contiguous blocks,

$$[\underbrace{\text{X-anc}}_{\rho_1} \mid \underbrace{H_a}_{n_a} \mid \underbrace{\text{ebit-TX}}_{c} \mid \underbrace{\text{ebit-RX}}_{c}], \qquad n_a = n - \rho_1 - c,$$

where the *X-anc* block carries the $\rho_1$ X-stabilizer pivots, the $H_a$ block holds the $k$ logical qubits and the $n_a - k$ Z-stabilizer "keep" ancillas, and the *ebit-TX/ebit-RX* blocks are the sender and receiver halves of the $c$ pre-shared Bell pairs. With this layout fixed:

- **$\mathcal{D}_0$:** reorder the $n_t$ qubits into the canonical block layout above (a column permutation of $H_{ex}$ and $H_{ez}$), then re-express the X-stabilizer generators so that the first $\rho_1$ columns of $H_{ex}$ form the identity matrix $I_{\rho_1}$. No Clifford gates are applied at this layer: the column permutation is absorbed into the qubit labeling, and the row reduction reflects a choice of stabilizer-generator basis;
- **$\mathcal{D}_1$:** apply CNOTs with controls in the X-anc block (column XORs on $H_{ex}$) that set the $H_a$-block columns of $H_{ex}$ to all-zero (eliminating the X-stabilizer's support on the $H_a$-block qubits) and bring its ebit-TX columns to match $D_x$; after this layer $H_{ex} = [\, I_{\rho_1} \mid 0 \mid D_x \mid D_x \,]$, so each ebit pair contributes the symmetric Bell-pair $XX$ stabilizer;
- **$\mathcal{D}_2$:** apply CNOTs (column XORs on $H_{ez}$) that set the last $\rho_1 - r_B$ X-anc-block columns of $H_{ez}$ to all-zero (eliminating the Z-stabilizer's support on the X-only-pivot qubits); the surviving rank-$r_B$ sub-matrix

in the first $r_B$ columns is the Z-pivot block $B_1$ carried by the "rank-deficient" X-anc qubits (those that hold both an X- and a Z-stabilizer pivot);

**$\mathcal{D}_3$:** apply CNOTs (column XORs on $H_{ez}$) that align the ebit-TX columns of $H_{ez}$ with $D_z$, so on each ebit pair the Z-stabilizer acts as the symmetric $ZZ$ Bell-pair stabilizer;

**$\mathcal{D}_4$:** apply CNOTs (column XORs on $H_{ez}$) that set the remaining $H_a$-block columns of $H_{ez}$ to all-zero (eliminating the Z-stabilizer's support there), using $B_1$ and the ebit-TX columns as control sources; after this layer $H_{ez} = [\, B_1 \mid 0 \mid 0 \mid D_z \mid D_z \,]$;

**$\mathcal{D}_5$:** apply Hadamards to the $\rho_1 - r_B$ X-anc qubits that carry an X-stabilizer pivot but no Z-stabilizer pivot. This swaps their $Z$ and $X$ stabilizers, so on the encoder side they are initialized as $|0\rangle$ (Z-stabilized) rather than $|+\rangle$ (X-stabilized).

After $\mathcal{D}$, every codeword factors as $|0\rangle^{\otimes(n-k-c)}$ on the Z-ancillas, $|\psi\rangle$ on the $k$ logical qubits, and $|\Phi^+\rangle^{\otimes c}$ on the ebit pairs. Running the six layers in reverse time order yields the encoder $\mathcal{E} = \mathcal{D}^\dagger$, which produces the encoded codeword from the same factored input.

# S3 QC-LDPC Parity-Check Matrices and Code Families

Let $P$ denote the $p \times p$ right circulant permutation matrix and $p$ be an odd prime. The parity-check matrices are constructed as arrays of circulant blocks:

$$H_x = \begin{bmatrix} P^{a_{0\cdot 0}} & P^{a_{0\cdot 1}} & \cdots & P^{a_{0\cdot(p-1)}} \\ \vdots & \vdots & \ddots & \vdots \\ P^{a_{(\ell_1-1)\cdot 0}} & \cdots & \cdots & P^{a_{(\ell_1-1)(p-1)}} \end{bmatrix}, \quad (1)$$

where each block row $i$ has exponents $\{k_i \cdot j \mod p : j = 0, \ldots, p-1\}$ for some $k_i \in \mathbb{F}_p$, and $\ell_1$ denotes the number of block rows. The matrix $H_z$ is constructed analogously with $\ell_2$ block rows using distinct exponent generators [2]. For the two-code construction (Theorem 1), we assume $\ell_1 = \ell_2$ and denote the per-matrix block row count as $\ell = \ell_1 = \ell_2$. Larger $\ell$ produces a denser encoder matrix. The code family optimized in the main paper is stated below.

**Theorem 1** (Two-code EA construction [2, Thm. 5])**.** *Let $C_1$ and $C_2$ be two classical QC-LDPC codes with parity-check matrices $H_x$ ($\ell_1$ block rows) and $H_z$ ($\ell_2$ block rows), respectively, where $1 \le \ell_1, \ell_2 \le p-1$ and $\ell_1 + \ell_2 \le p$, no block row is common to $H_x$ and $H_z$, and all block rows within each matrix are distinct. Then there exists an entanglement-assisted quantum QC-LDPC code with parameters $[[p^2,\ p^2 - 2p - (p-1)(\ell_1 + \ell_2 - 2) + 1;\ 1]]_2$ and girth greater than 4.*

**Theorem 2** (Two-code EA construction with $c = p-1$ [2, Thm. 6])**.** *Let $C_1$ and $C_2$ be two classical QC-LDPC codes whose parity-check matrices $H_x$ and $H_z$ consist of block rows with entries $P^{mj}$ over the block columns $j = 1, \ldots, p-1$, where the row multipliers $m$ are $1, \ldots, \ell_1$ for $H_x$ and $\ell_1+1, \ldots, \ell_1+\ell_2$ for $H_z$, with $2 \le \ell_1, \ell_2 \le p-2$ and $\ell_1 + \ell_2 \le p-1$. Then there exists an EA quantum QC-LDPC code with parameters $[[p^2 - p,\ p^2 - 3p - (p-1)(\ell_1 + \ell_2 - 2) + (p-1);\ p-1]]_2$, and the unassisted portion of its joint Tanner graph has girth greater than 4.*

**Theorem 3** (Single-code EA construction [2, Thm. 7])**.** *Let $C$ be a classical QC-LDPC code whose parity-check matrix $H$ contains $\ell \le p$ distinct block rows, where the block row with multiplier $k \in \mathbb{F}_p$ has entries $P^{kj}$ over the block columns $j = 0, \ldots, p-1$, and set $H_x = H_z = H$. Then there exists an EA quantum QC-LDPC code with parameters $[[p^2,\ (p-1)(p-\ell+1);\ p + (\ell-1)(p-1)]]_2$, and the Tanner graph of $C$ has girth greater than 4.*

The four instances evaluated in Table II of the main paper use the following multiplier sets. For Theorem 6, the code $[[42, 10; 6]]$ uses $H_x$ multipliers $\{1, 2, 3, 4\}$ and $H_z$ multipliers $\{5, 6\}$ at $p = 7$, and the code $[[42, 22; 6]]$ uses $\{1, 2\}$ and $\{3, 4\}$ at $p = 7$. The code $[[20, 10; 4]]$ uses $H_x$ multipliers $\{1, 2\}$ and $H_z$ multiplier

$\{3\}$ at $p = 5$. It has $\ell_2 = 1$, outside the stated hypotheses of Theorem 6, and its parameters and encoder are validated directly by the rank checks and the stabilizer-tableau test. For Theorem 7, the code $[[25, 16; 9]]$ uses $k \in \{1, 2\}$ at $p = 5$. The ebit columns $D_x$ and $D_z$ are obtained from a GF(2) factorization of $H_x H_z^T$ with inner dimension $c$. The published encoding scheme of [2] covers Theorems 5 and 6, and the same work notes that the argument extends to single-classical-code constructions. Its rank guarantee for Theorem 6 is verified there for a specific $H_z$ block-row choice, so every encoder used here is additionally checked by the stabilizer-tableau test. For the Theorem-7 instance, where $H_x = H_z$, the same layered construction emits a small number of degenerate self-target operations. These are removed, and the resulting encoder passes the stabilizer-tableau validation applied to every code in this work.

# S4 ASAP Two-Qubit Depth

The two-qubit depths quoted throughout this supplement and in Section IV of the main paper are obtained by the ASAP scheduler of Algorithm 1. The scheduler walks the forward CNOT list in its given time order and assigns each gate to the earliest time step in which both of its qubits are free. Two CNOTs are placed in the same time step if and only if their qubit sets are disjoint.

**Algorithm 1** ASAP two-qubit depth

**Require:** Ordered CNOT list $G = (g_1, \ldots, g_L)$ with $g_i = (c_i, t_i)$; qubit count $n$
**Ensure:** ASAP two-qubit depth $d$

ready$[q] \leftarrow 0$ for $q = 0, \ldots, n-1$
$d \leftarrow 0$
**for** $i = 1$ to $L$ **do**
  $s \leftarrow \max(\text{ready}[c_i], \text{ready}[t_i])$
  $e \leftarrow s + 1$
  ready$[c_i] \leftarrow e$;  ready$[t_i] \leftarrow e$
  $d \leftarrow \max(d, e)$
**end for**
**return** $d$

Running this procedure on the EAQC and beam CNOT lists of each code reproduces the depth values listed in this supplement.

# S5 Circuit Equivalence: A Four-Qubit Example

Different CNOT sequences can implement the same encoder matrix $M$; the shortest is what the optimization in the main paper targets. To illustrate, consider four qubits and the target $M = \begin{bmatrix} 1 & 0 & 0 & 1 \\ 0 & 1 & 0 & 1 \\ 0 & 0 & 1 & 1 \\ 0 & 0 & 0 & 1 \end{bmatrix}$. We compare two CNOT circuits (Fig. 1) and verify equivalence directly by tracking the cumulative matrix applied to $I_4$, step by step.

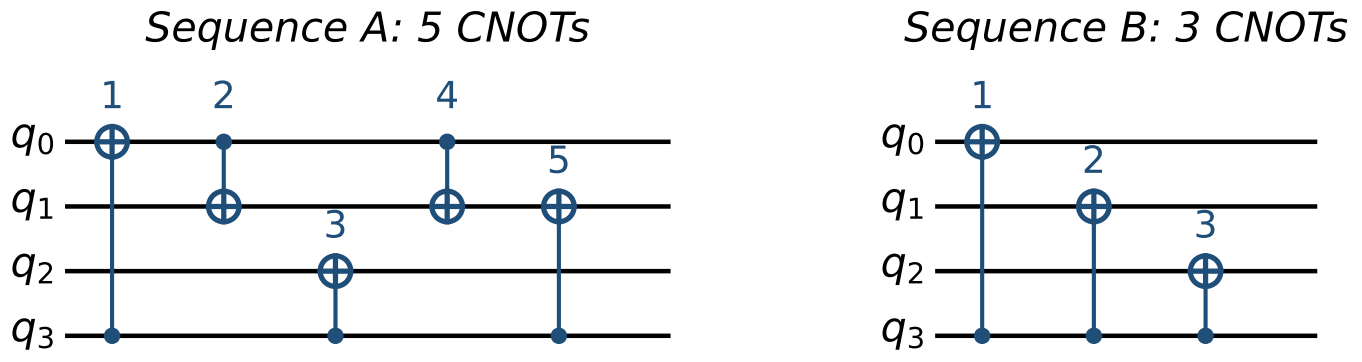


**Figure 1:** Two CNOT circuits implementing the same encoder matrix $M$. Sequence A (left) uses 5 gates; Sequence B (right) uses 3.

*Sequence A (5 CNOTs).* Starting from $M_0 = I_4$:

$$M_0 \xrightarrow{R_0 \leftarrow R_0 \oplus R_3} \begin{bmatrix} 1&0&0&1 \\ 0&1&0&0 \\ 0&0&1&0 \\ 0&0&0&1 \end{bmatrix} \xrightarrow{R_1 \leftarrow R_1 \oplus R_0} \begin{bmatrix} 1&0&0&1 \\ 1&1&0&1 \\ 0&0&1&0 \\ 0&0&0&1 \end{bmatrix} \xrightarrow{R_2 \leftarrow R_2 \oplus R_3} \begin{bmatrix} 1&0&0&1 \\ 1&1&0&1 \\ 0&0&1&1 \\ 0&0&0&1 \end{bmatrix}$$

$$\xrightarrow{R_1 \leftarrow R_1 \oplus R_0} \begin{bmatrix} 1&0&0&1 \\ 0&1&0&0 \\ 0&0&1&1 \\ 0&0&0&1 \end{bmatrix} \xrightarrow{R_1 \leftarrow R_1 \oplus R_3} \begin{bmatrix} 1&0&0&1 \\ 0&1&0&1 \\ 0&0&1&1 \\ 0&0&0&1 \end{bmatrix} = M.$$

*Sequence B (3 CNOTs).* Starting from $N_0 = I_4$:

$$N_0 \xrightarrow{R_0 \leftarrow R_0 \oplus R_3} \begin{bmatrix} 1&0&0&1 \\ 0&1&0&0 \\ 0&0&1&0 \\ 0&0&0&1 \end{bmatrix} \xrightarrow{R_1 \leftarrow R_1 \oplus R_3} \begin{bmatrix} 1&0&0&1 \\ 0&1&0&1 \\ 0&0&1&0 \\ 0&0&0&1 \end{bmatrix} \xrightarrow{R_2 \leftarrow R_2 \oplus R_3} \begin{bmatrix} 1&0&0&1 \\ 0&1&0&1 \\ 0&0&1&1 \\ 0&0&0&1 \end{bmatrix} = M.$$

The two final matrices are the same, so Sequence A and Sequence B implement the same row-XOR map on $\mathrm{GF}(2)^4$ and hence the same unitary on $(\mathbb{C}^2)^{\otimes 4}$—despite using different gate counts and different qubit pairs at the intermediate steps. The worked example below shows how the beam search finds such shorter realizations automatically.

# S6 Worked Example: Beam Search on the $[[6, 2; 2]]$ Encoder

This section traces the beam search of Algorithm 1 of the main paper on the smallest code in the test set, $[[6, 2; 2]]$, from the initial encoder matrix to the identity. All numbers below are produced by the same implementation and configuration used for Table III of the main paper (Hamming heuristic).

The tie-breaking and state-merging rules of the implementation are as follows. Ties are resolved deterministically. Candidates with equal $h(M')$ are ordered first by the number of operations applied so far and then by a fixed, seed-determined insertion order that follows the enumeration of the ordered pairs $(i, j)$. States that reach an identical matrix through different operation sequences are merged, and the earliest-generated representative with the fewest operations is retained. Repeated runs therefore produce identical results.

As an illustration, suppose $w = 2$ and three candidates share $h(M') = 9$. Candidate A was reached in 5 operations and generated third, candidate B in 5 operations and generated seventh, and candidate C in 6 operations and generated first. The first rule eliminates C, which uses the most operations, and the second rule ranks A before B by insertion order, so the beam retains A and B. The same two rules, applied in the same order, also select the representative kept when duplicate states are merged.

The baseline EAQC encoder for $[[6, 2; 2]]$ uses 11 CNOTs on $n_t = 8$ qubits. Its CNOT sub-sequence induces the binary matrix

$$M = \begin{bmatrix} 1&1&1&1&1&1&0&0 \\ 0&0&1&1&1&0&0&0 \\ 0&0&1&0&0&0&0&0 \\ 0&0&1&0&1&0&0&0 \\ 0&1&1&1&0&0&0&0 \\ 0&0&1&1&1&1&0&0 \\ 0&0&0&0&0&0&1&0 \\ 0&0&0&0&0&0&0&1 \end{bmatrix},$$

with $h(M) = \mathrm{wt}(M \oplus I) = 19$ off-identity entries. With beam width $w = 10$, the search reaches the identity in 9 row operations. Table S1 lists the sequence; each operation $R_i \leftarrow R_i \oplus R_j$ corresponds to one CNOT with control $j$ and target $i$, and applying the 9 operations in reverse order as CNOT gates realizes the CNOT sub-sequence of the optimized encoder.

**Table S1.** The 9-operation decomposition of $M$ found at $w = 10$. The column $h$ gives $\mathrm{wt}(M_t \oplus I)$ after step $t$; the column $\Delta$ gives the number of off-identity entries cleared by the step.

| Step $t$ | Operation | $h$ | $\Delta$ |
|---|---|---|---|
| 0 | – | 19 | – |
| 1 | $R_0 \leftarrow R_0 \oplus R_5$ | 15 | 4 |
| 2 | $R_3 \leftarrow R_3 \oplus R_1$ | 12 | 3 |
| 3 | $R_4 \leftarrow R_4 \oplus R_1$ | 9 | 3 |
| 4 | $R_5 \leftarrow R_5 \oplus R_1$ | 6 | 3 |
| 5 | $R_1 \leftarrow R_1 \oplus R_4$ | 4 | 2 |
| 6 | $R_1 \leftarrow R_1 \oplus R_2$ | 3 | 1 |
| 7 | $R_1 \leftarrow R_1 \oplus R_3$ | 2 | 1 |
| 8 | $R_0 \leftarrow R_0 \oplus R_1$ | 1 | 1 |
| 9 | $R_4 \leftarrow R_4 \oplus R_1$ | 0 | 1 |

The average clearing rate of this sequence is $19/9 \approx 2.11$ entries per operation, the value listed for $[[6, 2; 2]]$ in the clearing-rate table (Table S2) below.

*Why beam width matters here?* After step 1 the state is

$$M_1 = \begin{bmatrix} 1&1&0&0&0&0&0&0 \\ 0&0&1&1&1&0&0&0 \\ 0&0&1&0&0&0&0&0 \\ 0&0&1&0&1&0&0&0 \\ 0&1&1&1&0&0&0&0 \\ 0&0&1&1&1&1&0&0 \\ 0&0&0&0&0&0&1&0 \\ 0&0&0&0&0&0&0&1 \end{bmatrix}, \qquad h(M_1) = 15,$$

and exactly four operations tie at the minimum achievable next value $h = 12$:

$$R_1 \leftarrow R_1 \oplus R_4, \quad R_3 \leftarrow R_3 \oplus R_1, \quad R_4 \leftarrow R_4 \oplus R_1, \quad R_5 \leftarrow R_5 \oplus R_1.$$

A single-path search ($w = 1$) must commit to one of them; here it selects $R_1 \leftarrow R_1 \oplus R_4$ and reaches the identity in 10 operations, with the $h$-trace $19, 15, 12, 10, 8, 5, 4, 3, 2, 1, 0$. The $w = 10$ beam instead keeps all four tied states. The branch that applied $R_3 \leftarrow R_3 \oplus R_1$ turns out to admit three consecutive 3-clearing operations (steps 2–4 of Table S1, $15 \to 12 \to 9 \to 6$) and finishes in 9 operations, with the $h$-trace $19, 15, 12, 9, 6, 4, 3, 2, 1, 0$. The two branches are indistinguishable to the heuristic at the moment of the tie; the advantage of the retained branch appears only at step 3, where the two $h$-traces first diverge (10 versus 9). This is the mechanism behind the $w = 1$ versus $w = 10$ entries for $[[6, 2; 2]]$ in Table III of the main paper (10 versus 9 CNOTs) and, more generally, behind the gap between the greedy baseline and beam search in Table II of the main paper.

The worked example above uses the code $[[6, 2; 2]]$ of [1], specified by

$$H_x = \begin{bmatrix} 0&0&1&0&1&0 \\ 1&0&0&0&0&1 \\ 0&1&0&1&0&0 \end{bmatrix}, \quad H_z = \begin{bmatrix} 0&1&0&0&0&1 \\ 0&0&1&1&0&0 \\ 1&0&0&0&1&0 \end{bmatrix},$$

$$D_x = \begin{bmatrix} 1&0 \\ 0&1 \\ 1&1 \end{bmatrix}, \quad D_z = \begin{bmatrix} 0&1 \\ 1&0 \\ 1&1 \end{bmatrix}.$$

For completeness, the second benchmark code $[[8, 2; 2]]$, constructed for this study, is specified by

$$H_x = \begin{bmatrix} 0&0&0&1&0&1&1&0 \\ 0&1&0&0&1&1&0&0 \\ 1&1&0&1&0&1&1&1 \\ 0&1&1&1&1&1&0&0 \end{bmatrix}, \quad H_z = \begin{bmatrix} 1&1&1&1&1&1&1&1 \\ 1&1&1&1&1&0&0&1 \\ 1&1&0&0&1&0&0&0 \\ 0&1&0&0&0&0&1&0 \end{bmatrix},$$

$$D_x = \begin{bmatrix} 1&1 \\ 1&0 \\ 0&0 \\ 1&0 \end{bmatrix}, \quad D_z = \begin{bmatrix} 1&0 \\ 0&1 \\ 0&0 \\ 1&0 \end{bmatrix}.$$

# S7 Effect of Block Row Count and Clearing Rate

Table II of the main paper shows that the number of block rows $\ell$ is the dominant determinant of optimization potential for the tested codes. For $\ell = 1$ codes the reductions are modest (7.3%–23.5%), with the smallest reduction on the largest code $[[121, 100; 1]]$. For the Theorem-5 codes with $\ell \geq 2$, the reduction is more substantial (31.5%–34.0%). This difference is structural: when $\ell = 1$, the encoder matrix $M$ has dispersed

off-diagonal support, so most row operations remove at most one mismatch with the identity. When $\ell \geq 2$, the denser block structure creates correlated, overlapping rows that allow one row operation to eliminate multiple terms.

This effect can be quantified. Define the *clearing rate* of a decomposition of $M$ with $L$ CNOTs as $h(M)/L$, where $h(M) = \mathrm{wt}(M \oplus I)$: the average number of off-identity entries removed per CNOT, with worst case one. Table S2 compares the EAQC sequence with the beam result. For the Theorem-5 $\ell \geq 2$ codes the EAQC sequence clears almost one entry per CNOT (1.04–1.13): its fixed layer order cannot use the overlap that the dense blocks create. Beam search lifts the rate to about 1.6, and that headroom is the 31–34% reduction. For the $\ell = 1$ codes the EAQC sequence is already efficient (1.43–1.71), so less headroom remains. For the Theorem-6/7 codes, the EAQC sequences start from higher clearing rates (1.33–2.19) and beam search again lifts them (1.94–2.78), with reductions of 12.8–33.0% (Table II).

**Table S2.** Off-identity weight $h(M)$ and clearing rates for the twelve tested codes.

| **Code** | $\ell$ | $h(M)$ | $h/L_{\mathrm{EAQC}}$ | $h/L_{\mathrm{Beam}}$ | **Red.** |
|---|---|---|---|---|---|
| $[[9, 4; 1]]$ | 1 | 29 | 1.71 | 2.23 | 23.5% |
| $[[25, 16; 1]]$ | 1 | 55 | 1.45 | 1.67 | 13.2% |
| $[[25, 8; 1]]$ | 2 | 101 | 1.13 | 1.66 | 31.5% |
| $[[49, 36; 1]]$ | 1 | 117 | 1.43 | 1.58 | 9.8% |
| $[[49, 12; 1]]$ | 3 | 352 | 1.04 | 1.58 | 34.0% |
| $[[121, 100; 1]]$ | 1 | 313 | 1.44 | 1.55 | 7.3% |
| $[[6, 2; 2]]$ | – | 19 | 1.73 | 2.11 | 18.2% |
| $[[8, 2; 2]]$ | – | 17 | 1.31 | 1.70 | 23.1% |
| $[[20, 10; 4]]$ | 3 | 103 | 2.19 | 2.78 | 21.3% |
| $[[25, 16; 9]]$ | 2 | 302 | 1.69 | 1.94 | 12.8% |
| $[[42, 10; 6]]$ | 6 | 375 | 1.33 | 1.95 | 31.7% |
| $[[42, 22; 6]]$ | 4 | 661 | 1.80 | 2.69 | 33.0% |

# S8 Code $[[25, 8; 1]]$

The encoder for $[[25, 8; 1]]$ acts on $n_t = 26$ qubits. The baseline EAQC encoder uses 8 Hadamards and 89 CNOTs at ASAP depth 26 (Fig. 2); beam search reduces the CNOT count to 61 and the depth to 22 (Fig. 3). The reductions are 31% in count and 15% in depth.

*Qubit-role partition for* $[[25, 8; 1]]$*.* The EAQC column permutation places the qubits in this order: $q_0$ is the rank-$r_B$ Z-stabilizer pivot; $q_1$–$q_8$ are X-stabilizer ancillas (which receive the eight fixed Hadamards); $q_9$–$q_{15}$ are the seven Z-stabilizer "keep" ancillas in the $\mathrm{H}_a$ block; $q_{16}$–$q_{23}$ are the eight logical qubits, initialized as $|\psi_1\rangle, \ldots, |\psi_8\rangle$; $q_{24}$ holds the sender's half of the shared ebit pair; and $q_{25}$ holds the receiver's half (untouched by the encoder).

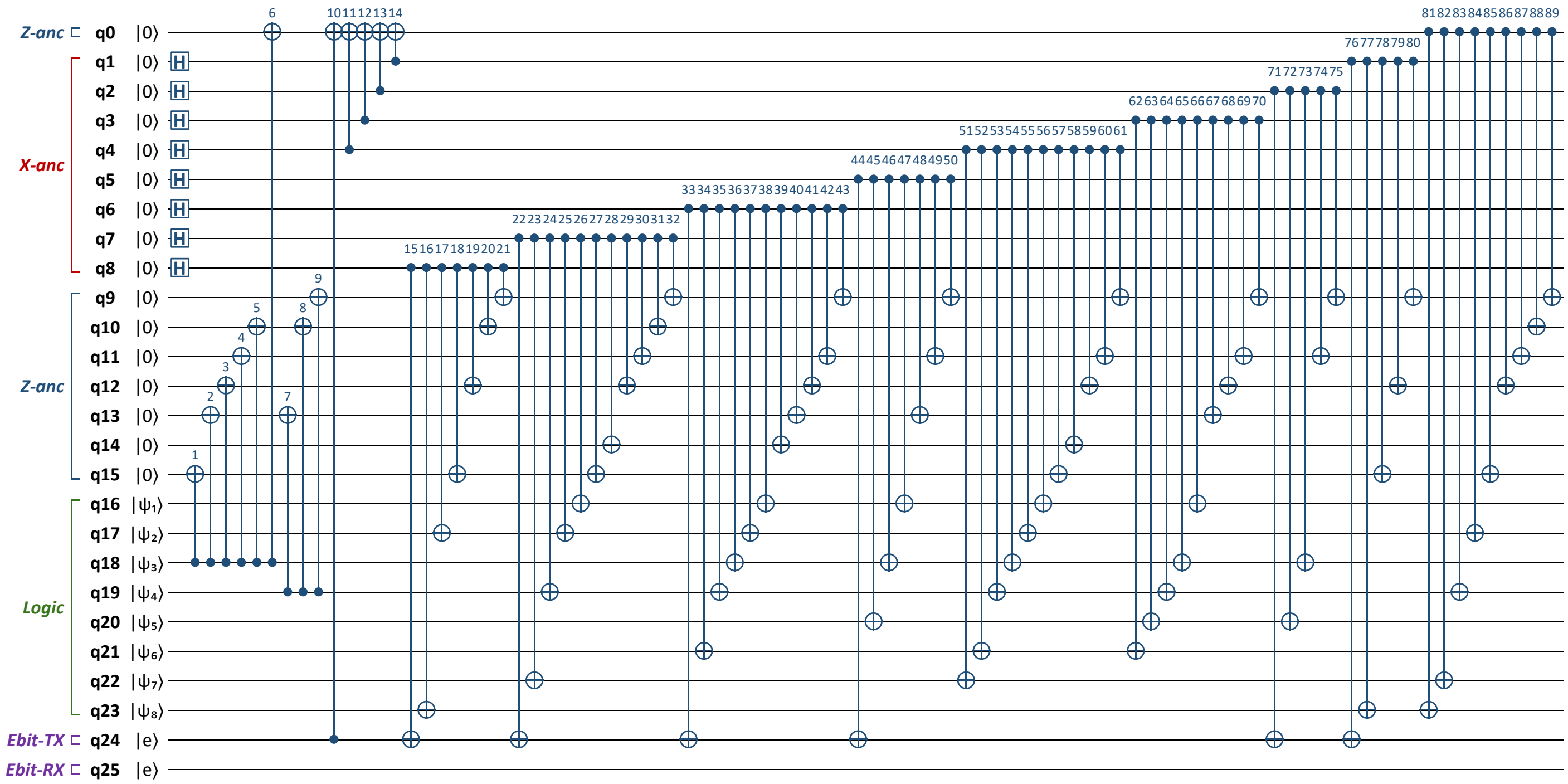


**Figure 2:** Baseline EAQC encoder for [[25, 8; 1]]: 8 Hadamards and 89 CNOTs (ASAP two-qubit depth 26).

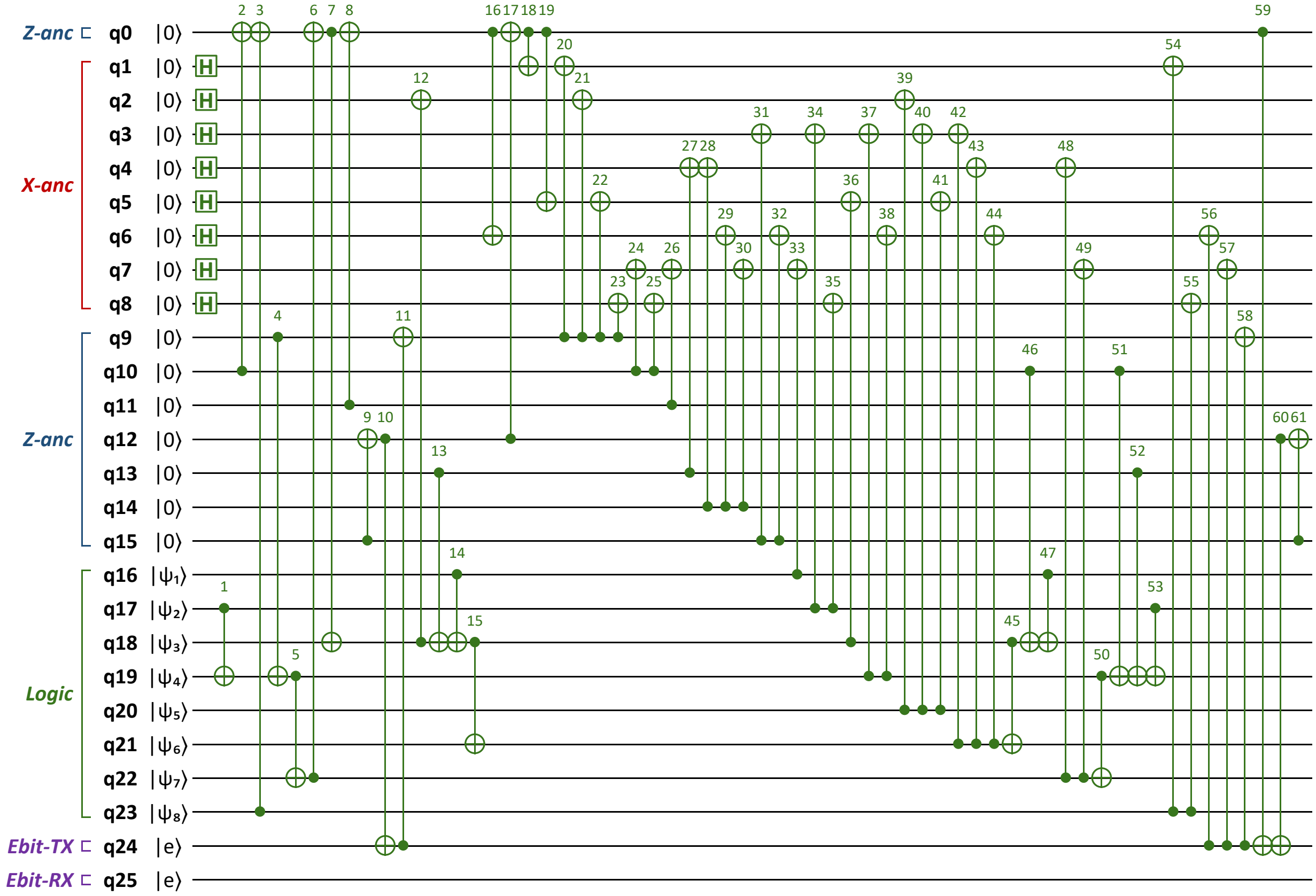


**Figure 3:** Beam-search-optimized encoder for $[[25, 8; 1]]$: 8 Hadamards and 61 CNOTs (ASAP two-qubit depth 22).


\end{document}